# ON THE WAVY SYMMETRY OF THE SOLAR SYSTEM PLANETARY AND SATELLITES DISTANCES SCALES

© Garry L. Abramyan

The Nizhny Novgorod State Technical University
Nizhny Novgorod, Russia
E – mail: esvm @ nntu.sci-nnov.ru
Home/ 603006 Gorky Str., H 149A, Apt 18, Nizhny Novgorod, Russia
Tel. (+007) ( 8312 ) 192-593

The symmetry of the undulating spacial distributions of the differences $r_{k+1} - r_k$ and $r_{k,i+1} - r_{k,i}$ of the dimensionless $r$ - parameters $r_k = CT_k / (2\pi a_k)$ and $r_{k,i} = CT_{k,i} / (2\pi a_{k,i})$ in dependence on the numbers $k$ and $i$ for the orbits of the Solar system planets, satellites and rings of the Jupiter and Saturn is found. Here $a_k$ is the major half axis of the planet's orbit and $T_k$ is the sidereal period rotation of the $k$ th planet around the Sun and $a_{k,i}$ is the major half axis of orbit, $T_{k,i}$ is the sidereal period rotation of the $i$ th satellites or rings of the $k$ th planet, $k$ and $i$ count off one after another with increase of the distances $a_k$ and $a_{k,i}$, $C$ is the group velocity of light. The global dependence of planetary distances scale $a_k$ and satellites and rings of the planets distances scale $a_{k,i}$ on the sidereal period rotation and radius of the Sun has been found.

## 1. INTRODUCTION

In this paper we analyze the solar system functional connections ensuring fulfillment of the Titius – Bode Law of Planetary Distances scale. All attempts to explain this law within the framework of known formal physical systems ended in failure ( Nieto, 1972). That is why our attention have been directed toward the investigation of the experimental astronomical data which are not some direct logic consequences of the known axiomatic of some physical theory. The method of the investigation is based on the analysis of spatial distributions of differences of the some dimensionless r – parameter ; it is composed with magnitudes that described Kepler's orbit of planets and their satellites and the group velocity of light. Most of the results derived in Section 2.2 and 2.3 will be shown and thus there is no need for us to repeat the detailed analysis of these paragraphs here. It gives astronomically interesting picture of the wavy, undulating symmetry of the Solar system.

## 2. DATA ANALYSIS

### 2.1. General Formulation

The Sun surface gravity in its equatorial zone $g_{\otimes e}$ obeys the experimental relation feasible with relative error of 0,2% (Abramyan, 2001).

$$g_{\otimes e} = g(B) = \gamma \frac{M_\otimes}{R_\otimes^2(B)} = \frac{2C}{T_\otimes(B)}, \quad -16^\circ \le B \le +16^\circ \qquad (1)$$

In (1) $C$ is the group velocity of light, $g_{\otimes e}$ is the gravity on the surface of the Sun, function $T_\otimes(B)$ is the sidereal period rotation of the Sun and function $R_\otimes(B)$ is the radius of the Sun corresponding with heliographic latitude $B$, $\gamma$ is the gravity physical constant, $M_{\otimes}$ is the mass of the Sun.

The experimental relation (1) testifies to dependence $\gamma$ upon the sidereal period rotation of the substance in the solar system. That is why the Third Kepler Law in Newtonian form can be represented otherwise (Abramyan, 2001):

$$a_k^3 = \frac{g_{\otimes e}}{4\pi^2} R_{\otimes e}^2 T_k^2 = \frac{C R_{\otimes e}^2 T_k^2}{2\pi^2 T_{\otimes e}} \quad , \tag{2}$$

In equality (2) $a_k$ is the major half axis of elliptical orbit of $k$ th planet, $T_k$ is the sidereal period rotation of $k$ th planet around the Sun, $k = 1,2,...,9$ from $k = 1$ for Mercury to Pluto with $k = 9$ ,

$$R_{\otimes e} \equiv R_{\otimes}(B), \qquad T_{\otimes e} \equiv T_{\otimes}(B), \qquad -16^{\circ} \le B \le +16^{\circ} \ . \tag{3}$$

Expression (2) has no contradictions to the presentation of the Third Kepler Law in Newtonian form since the mass of the Sun is more greater than the mass of all planets. The presentation of the Third Kepler Law in the form of equation (2) indicates the opportunity to obtain the Sun surface gravity in its equatorial zone by means of measurements of the Sun radius and the major half axis of elliptical orbit and the sidereal period rotation of some planet around the Sun.

By means of transformation (2) we determine $r$ - parameter as relation:

$$r_k \equiv \frac{CT_k}{2\pi a_k} = \frac{C}{R_{\otimes e}\sqrt{g_{\otimes e}}}\sqrt{a_k} = \frac{1}{R_{\otimes e}}\sqrt{\frac{CT_{\otimes e}}{2}}\sqrt{a_k} \ . \tag{4}$$

The $r$ - parameter $r_k$ submitted for consideration in (4) may be presented as a ratio of group velocity of light $C$ to the mean orbital velocity of the $k$ th planet $2\pi a_k / T_k$ .

The computation of quantities $r_k$ for nine large planets of the solar system gives striking behavior of differences $r_{k+1} - r_k$ with a number $k$ for $k = 1,2,...,8$ .

## 2.2. Planetary distances scale.

Direct computations $r_{k+1} - r_k$ from planet to planet with world-wide known parameters $T_k$ and $a_k$ for Earth group of planets in matrix notation yields:

$$\begin{bmatrix} r_2 - r_1 \\ r_3 - r_2 \\ r_4 - r_3 \end{bmatrix} = \begin{bmatrix} 2.292 \\ 1.508 \\ 2.347 \end{bmatrix} \cdot 10^3 \ , \tag{5}$$

Calculations $r_{k+1} - r_k$ for outer group of planets behind the belt of asteroids gives:

$$\begin{bmatrix} r_6 - r_5 \\ r_7 - r_6 \\ r_8 - r_7 \\ r_9 - r_8 \end{bmatrix} = \begin{bmatrix} 8.090 \\ 12.950 \\ 11.080 \\ 8.085 \end{bmatrix} \cdot 10^3 , \tag{6}$$

It is important to note the change of $r$ - parameter $r_{k+1} - r_k$ appears wavy dependence and symmetry upon number $k$ for boundary planets of both groups of planet independently of $a_k$, mass and size of planets

$$r_2 - r_1 \approx r_4 - r_3 \;, \quad r_6 - r_5 \approx r_9 - r_8 \,. \tag{7}$$

The ratio of the normalizer $N_2$ for the distribution of the $r$ - parameter differences of the outer group of planet to the normalizer $N_1$ for the distribution of the $r$ - parameter differences of the Earth group of planet is approximately $N_2 / N_1 \cong 2\pi$.

$$N_1 = \sum_{k=1}^{3} (r_{k+1} - r_k) = 6.15 \cdot 10^3 \approx 2\pi\, 10^3 \;, \tag{8}$$

$$N_2 = \sum_{k=5}^{8} (r_{k+1} - r_k) = 40.20 \cdot 10^3 \approx (2\pi)^2\, 10^3 \,. \tag{9}$$

We attempted to approximate differences $r_{k+1} - r_k$ from (5) and (6) and obtained recurrence relations:

$$\begin{aligned} r_{k+1} - r_k \approx 12\pi[ & 3\cdot 5(k-2)^2(k-3)^2 + \\ & + 2\cdot 4\cdot 5(k-1)^2(k-3)^2 + \\ & + 3\cdot 5(k-1)^2(k-2)^2 \,], \quad k = 1,2,3; \end{aligned} \tag{10}$$

$$\begin{aligned} r_{k+1} - r_k \approx 14\pi[ & 5(k-6)^2(k-7)^2(k-8)^2 + \\ & + 7\cdot 9(k-5)^2(k-6)^2(k-8)^2 + \\ & + 8\cdot 9(k-5)^2(k-7)^2(k-8)^2 + \\ & + 5(k-5)^2(k-6)^2(k-7)^2] \;, \; k = 5,6,7,8. \end{aligned} \tag{11}$$

Direct computations $r_{k+1} - r_k$ are in good agreement with equations (10) and (11). It is not difficult to predict scale of distance $a_k$ using (4),(10) and (11) in both Earth and outer groups of planets that leads to the Titius – Bode Law represented with recurrence relations:

$$a_{k+1} = \left[ \sqrt{a_k} + \frac{\sqrt{2}\, R_{\otimes e}}{\sqrt{CT_{\otimes e}}} (r_{k+1} - r_k) \right]^2 \,. \tag{12}$$

Here in (12) $k = 1, 2,3,5,6,7,8$;

$$a_5 = \left[ \sqrt{a_A} + \frac{\sqrt{2}\, R_{\otimes e}}{\sqrt{CT_{\otimes e}}} (r_5 - r_A) \right]^2 ; \tag{13}$$

$$a_A = \left[ \sqrt{a_4} + \frac{\sqrt{2}\, R_{\otimes e}}{\sqrt{CT_{\otimes e}}} (r_A - r_4) \right]^2 . \tag{14}$$

Here in (13) and (14) $r_A$ is the normalized $r$ - parameter of the belt of asteroids:

$$r_A = \frac{r_4 + r_5}{2} \quad (15)$$

and $a_A$ is the “radius” of the belt of asteroids.

Expressions (12),(10) and (11) describe the Titius-Bode Law of Planetary Distances scale with relative error of 1% and remove all doubts and complications connected with its mathematical description ( Nieto, 1972).

Numerical analysis showed the equation (12) approximately represents geometric progression

$$a_{k+1} = q_k\, a_k\ . \quad (16)$$

Here “denominator” $q_k$ is not a constant but sluggish and slows function of number $k$ :

$$q_k = \left[1 + \frac{1}{\sqrt{a_k}} \frac{\sqrt{2}R_{\otimes e}}{\sqrt{CT_{\otimes e}}}(r_{k+1} - r_k)\right]^2 \equiv$$

$$\equiv [1 + (r_{k+1} - r_k)/r_k]^2 ; \quad (17)$$

$$1.3 \le q_k \le 2\ . \quad (18)$$

This property $q_k$ from (18) allows us to use $q_{k-1} \approx q_k$ for rough estimation $a_{k+1}$ with approximate relation $a_{k+1} \approx q_{k-1}a_k$ .The estimation $a_{k+1}$ can be used for estimation $r_{k+1} - r_k$ and further making more precise $a_{k+1}$.

2.3.Planets satellites distances scale.

We shall take multisatellites systems of the Jupiter and the Saturn to display the results of the investigation of the physical functional connections ensuring fulfillment of the Titius – Bode Law for satellites and rings of planets.

Let us represent the Third Kepler Law for $i$ th satellite of the $k$ th planet by analogy with expression (2) in the form of equation

$$a_{k,i}^3 = \frac{g_k}{4\pi^2} R_k^2 T_{k,i}^2\ . \quad (19)$$

Here $a_{k,i}$ is the major half axis of elliptical orbit, $T_{k,i}$ is the sidereal period rotation of $i$ th satellite or ring of $k$ th planet, $g_k$ is $k$ th planet surface gravity. Number $i$ count off one after another with increase of distances $a_{k,i}$ .

Let us introduce by analogy with experimental relation (1) ”parameter rotation” $T_{\otimes k}$ of $k$ th planet by means of substitution of $T_{\otimes k}$ for $g_k$ :

$$g_k = \gamma \frac{m_k}{R_k^2} \equiv \frac{2C}{T_{\otimes k}}\ . \quad (20)$$

Here $m_k$ is a mass, $R_k$ is a radius of the $k$ th planet.

From expression (19) and (20) it follows that $r$ - parameter for satellite system may be determined by relation:

$$r_{k,i} \equiv \frac{CT_{k,i}}{2\pi a_{k,i}} = \frac{C}{R_k\sqrt{g_k}}\sqrt{a_{k,i}} \equiv \frac{1}{R_k}\sqrt{\frac{CT_{\otimes k}}{2}}\sqrt{a_{k,i}} \quad . \tag{21}$$

From expressions (20) and (21) it follows that "parameter rotation" $T_{\otimes k}$ may be represented as substitution of $T_{\otimes k}$ for $g_k$:

$$T_{\otimes k} \equiv \frac{2C}{g_k} \tag{22}$$

and the other way as relation

$$T_{\otimes k} = \frac{2R_k^2}{C}\frac{r_{k,i}^2}{a_{k,i}}. \tag{23}$$

The computations of quantities $T_{\otimes k}$ with (22) и (23) coincide with relative error of 1%.

Notice that the direct computation of all differences $r_{k,i+1} - r_{k,i}$ with expression $r_{k,i} = CT_{k,i} / (2\pi a_{k,i})$ from (21) is complicated because of absence of reliable meaning of $T_{k,i}$.In this case we may calculate $T_{\otimes k}$ with (22) and with the right part of (21)we can find all $r_{k,i}$ .From left part of (21) it follows that

$$T_{k,i} = \frac{2\pi a_{k,i} r_{k,i}}{C} \quad . \tag{24}$$

The procedure just described can be used in practical calculations of the periods rotation $T_{6,1}$ , $T_{6,2}$ , $T_{6,3}$ and $T_{6,6}$ of the Saturn ring $C, B, A$ and $F$ accordingly. The computation $T_{6,i}$ for rings of Saturn gives:

$$T_{6,1} = 0.27907T_* \ ; \ T_{6,2} = 0.39983T_* \ ;$$
$$T_{6,3} = 0.54318T_* \ ; \ T_{6,4} = 0.62000T_* \, . \tag{25}$$

Here $T_*$ is the sidereal twenty four hours.

From expression (21) there can be obtained the relation that describes distance scale $a_{k,i}$ in the form analogous with (12):

$$a_{k,i+1} = \left[\sqrt{a_{k,i+1}} + \beta_k\left(r_{k,i+1} - r_{k,i}\right)\right]^2 \ , \tag{26}$$

Here

$$\beta_k \equiv \frac{\sqrt{2}R_k}{\sqrt{CT_{\otimes k}}} \equiv \frac{\sqrt{a_{k,i+1}} - \sqrt{a_{k,i}}}{r_{k,i+1} - r_{k,i}} \ , \tag{27}$$

$k = 1, 2, ..., 9$ , $i$ is a number of satellite or of ring of the $k$ th planet.

The numerical analysis shows that the parameters $\beta_k$ for system of satellites and rings of Jupiter and Saturn are:

$$\beta_5 = 11{,}87\, ; \quad \beta_6 = 6{,}5313 \ . \tag{28}$$

### 2.3.1 Saturn's distances scale

The analysis of the spatial distribution $r_{6,i+1} - r_{6,i}$ was fulfilled for all nowadays known twenty seven satellites and four rings of the Saturn including twelve its remote satellites discovered in 2000 (Marsden, 2000 a, b, c, d; see Minor Planet Electronic Circular 2000-Y13,14,15,33 by address:< http://cfa-www.harvard.edu/iau/mpc.html >). Direct computation $r_{6,i+1} - r_{6,i}$ in matrix notation gives:

$$\begin{bmatrix} r_{6,2} - r_{6,1} \\ r_{6,3} - r_{6,2} \end{bmatrix} = \begin{bmatrix} 1.771 \\ 1.685 \end{bmatrix} \cdot 10^3 \ , \tag{29}$$

$$\begin{bmatrix} r_{6,4} - r_{6,3} \\ r_{6,5} - r_{6,4} \\ r_{6,6} - r_{6,5} \\ r_{6,7} - r_{6,6} \\ r_{6,8} - r_{6,7} \end{bmatrix} = \begin{bmatrix} 0.578 \\ 0.137 \\ 0.078 \\ 0.071 \\ 0.613 \end{bmatrix} \cdot 10^3 \ , \tag{30}$$

$$\begin{bmatrix} r_{6,9} - r_{6,8} \\ r_{6,10} - r_{6,9} \\ r_{6,11} - r_{6,10} \\ r_{6,12} - r_{6,11} \\ r_{6,13} - r_{6,12} \\ r_{6,14} - r_{6,13} \end{bmatrix} = \begin{bmatrix} 0.003 \\ 2.162 \\ 2.725 \\ 2.639 \\ 3.423 \\ 0.016 \end{bmatrix} \cdot 10^3 , \tag{31}$$

$$\begin{bmatrix} r_{6,15} - r_{6,14} \\ r_{6,16} - r_{6,15} \\ r_{6,17} - r_{6,16} \end{bmatrix} = \begin{bmatrix} 5.368 \\ 18.325 \\ 5.842 \end{bmatrix} \cdot 10^3 , \tag{32}$$

$$\begin{bmatrix} r_{6,18} - r_{6,17} \\ r_{6,19} - r_{6,18} \end{bmatrix} = \begin{bmatrix} 32.003 \\ 71.53 \end{bmatrix} \cdot 10^3 , \tag{33}$$

$$\begin{bmatrix} r_{6,20} - r_{6,19} \\ r_{6,21} - r_{6,20} \\ r_{6,22} - r_{6,21} \\ r_{6,23} - r_{6,22} \\ r_{6,24} - r_{6,23} \\ r_{6,25} - r_{6,24} \\ r_{6,26} - r_{6,25} \end{bmatrix} = \begin{bmatrix} 0.46 \\ 10.96 \\ 13.83 \\ 3.59 \\ 5.15 \\ 7.27 \\ 0.47 \end{bmatrix} \cdot 10^3 , \tag{34}$$

$$\begin{bmatrix} r_{6,27}-r_{6,26} \\ r_{6,28}-r_{6,27} \\ r_{6,29}-r_{6,28} \\ r_{6,30}-r_{6,29} \\ r_{6,31}-r_{6,30} \end{bmatrix} = \begin{bmatrix} 0.12 \\ 2.82 \\ 1.23 \\ 6.64 \\ 16.19 \end{bmatrix} \cdot 10^3 \ . \qquad (35)$$

The spatial distribution $r_{6,i+1} - r_{6,i}$ in the satellites and rings system of the Saturn appears wavy undulating dependence and symmetry upon number $i$ independently of the size $R_k$ ,mass $m_k$ and quantity $a_{6,i}$ in every group with the numbers $i = 1,2$ from (29), $i = 3,4,5,6,7$ from (30), $i = 8,9,10,11,12,13$ from (31), $i = 14,15,16$ from (32), $i = 19,20,21,23,24,25$ from (34),except groups of satellites with the numbers $i = 17,18$ from (33) and $i = 26,27,28,29,30$ from (35) :

$$r_{6,2} - r_{6,1} \approx r_{6,3} - r_{6,2} \, ;$$
$$r_{6,4} - r_{6,3} \approx r_{6,8} - r_{6,7} \, ;$$
$$r_{6,9} - r_{6,8} \approx r_{6,14} - r_{6,13} \, ;$$
$$r_{6,15} - r_{6,14} \approx r_{6,17} - r_{6,16} \, ;$$
$$r_{6,20} - r_{6,19} \approx r_{6,26} - r_{6,25} \, . \qquad (36)$$

2.3.2 Jupiter’s distances scale

Direct computation $r_{5,i+1} - r_{5,i}$ for satellites and ring of Jupiter in matrix notation gives:

$$\begin{bmatrix} r_{5,2}-r_{5,1} \\ r_{5,3}-r_{5,2} \end{bmatrix} = \begin{bmatrix} 0.08 \\ 0.02 \end{bmatrix} \cdot 10^3 , \qquad (37)$$

$$\begin{bmatrix} r_{5,4}-r_{5,3} \\ r_{5,5}-r_{5,4} \\ r_{5,6}-r_{5,5} \\ r_{5,7}-r_{5,6} \end{bmatrix} = \begin{bmatrix} 1.85 \\ 1.12 \\ 4.78 \\ 4.52 \end{bmatrix} \cdot 10^3 , \qquad (38)$$

$$\begin{bmatrix} r_{5,8}-r_{5,7} \\ r_{5,9}-r_{5,8} \\ r_{5,10}-r_{5,9} \\ r_{5,11}-r_{5,10} \end{bmatrix} = \begin{bmatrix} 5.74 \\ 8.97 \\ 52.28 \\ 1.43 \end{bmatrix} \cdot 10^3 , \qquad (39)$$

$$\begin{bmatrix} r_{5,12}-r_{5,11} \\ r_{5,13}-r_{5,12} \\ r_{5,14}-r_{5,13} \\ r_{5,15}-r_{5,14} \end{bmatrix} = \begin{bmatrix} 0.94 \\ 0.12 \\ 29.94 \\ 4.74 \end{bmatrix} \cdot 10^3 , \qquad (40)$$

$$\begin{bmatrix} r_{5,16}-r_{5,15} \\ r_{5,17}-r_{5,16} \end{bmatrix} = \begin{bmatrix} 2.65 \\ 11.62 \end{bmatrix} \cdot 10^3 . \qquad (41)$$

Notice that spatial distribution $r_{5,i+1}-r_{5,i}$ in satellites system of the Jupiter appears wavy undulating dependency and symmetry upon number $i$ as in the case of the Saturn system of satellites.

## 3. THE INTERRELATION OF THE PLANETARY AND SATELLITES DISTANCES SCALES

The most astronomically interesting fact is that not only planetary distances scale depends on the size and sidereal period rotation of the Sun. The expression for the planet surface gravity dependence $g_k$ on the $g_{\otimes e}$, sidereal period rotation of planet around the Sun $T_k$ and the effective radius of planet $R_k$ has been obtained (Abramyan, 2001):

$$g_k = \gamma \frac{m_k}{R_k^2} \equiv \frac{4\pi R_k}{T_k^2} \left( \frac{M_\otimes}{m_k} \right)^{\alpha_k} =$$

$$= (4\pi)^{\frac{1}{1+\alpha_k}} g_{\otimes e}^{\frac{\alpha_k}{1+\alpha_k}} \frac{1}{T_k^{\frac{2}{1+\alpha_k}}} \frac{R_\otimes^{\frac{2\alpha_k}{1+\alpha_k}}}{R_k^{\frac{2\alpha_k - 1}{1+\alpha_k}}} \quad ; \qquad (42)$$

$$\sum_{k=1}^{9} \frac{2C}{T_k} = \sum_{k=1}^{9} g_k = \frac{g_{\otimes e}}{2} \equiv \frac{C}{T_{\otimes e}} \ . \qquad (43)$$

The equation (42) can be solved for the unknowns $\alpha_k$.The nine parameters $\alpha_k$ that satisfy the equation (42) may be expressed in matrix notation:

$$\begin{bmatrix} \alpha_1 \\ \alpha_2 \\ \alpha_3 \\ \alpha_4 \\ \alpha_5 \\ \alpha_6 \\ \alpha_7 \\ \alpha_8 \\ \alpha_9 \end{bmatrix} = \begin{bmatrix} 1.0016298 \\ 1.3571473 \\ 1.4645268 \\ 1.3018916 \\ 3.1810311 \\ 2.8505207 \\ 2.6048175 \\ 2.9685156 \\ 3.1804322 \end{bmatrix} . \qquad (44)$$

It should be taken into notice that $\alpha_5 \cong \alpha_9$ at conditions of distinction in gravity on the surface of Jupiter and Pluto and matrix equality (44) approximates with symmetric expressions:

$$\alpha_k = -0.11572\,(k-3)^2 + 1.4645268\,, \quad k = 1,2,3,4\,; \qquad (45)$$

$$\alpha_k = +0.14405\,(k-7)^2 + 2.6048175\,, \quad k = 5,6,7,8,9\,. \qquad (46)$$

## 4. CONCLUSIONS

As previously observed the gravity physical constant $\gamma$ depends on sidereal period rotation of the substance in accordance with experimental relation (1) feasible with relative error of 0,2 %. It may be supposed that rotation of bodies is essential property of gravity as well as the gravity property to penetrate through any screens to another rotating bodies independently of their chemical properties. Taking into account (1),(2),(19) and the expressions (42) and (43) for the planet surface gravity $g_k$ dependence on the $g_{\otimes e}$, it is concluded that period rotation of planets around the Sun and size of planets and planetary and satellites distances scales in solar system connected with rotation period and the Sun size. It is evident that the Sun gravity dominates in solar system and changes in its rotation, especially during its active phase, and it can cause global variability of planetary distances scale, change altitudes of the satellites and rings and radius of planets and may cause earthquakes. It may be interesting to investigate the possible correlation between solar bursts and changing distances between planets and their satellites.

## 5. DISCUSSION

The physical laws are idealizations of reality. There are difference between formal mathematical systems in physics and reality of the Universe. We are able to measure seeming angles between astronomical objects but we use some formal virtual model of geometry of the world for measuring distances to remote sources. There are no instruments capable to measure time continuously and we count time with cycle timer. There are no continuity of time for us, but we generally use the continuous variable for time.

As our knowledge grows we observe that a given physical situation can be idealized mathematically in a number of different ways. It is therefore necessary to choose new unifying theory that remove defects of old one and possesses properties of completeness and uniqueness description of all discovered natural phenomena.

It is evident the absence of uniqueness solutions of classical mechanics with its time symmetry. It is well known incompleteness of the classical mechanics for the solution of the N-body motion problem with intrinsic gravity. In fact, the motion of mass point is described in Newtonian mechanics with second-order differential equations. Thus, the solution of the N-body motion problem in three- dimensional coordinates space of 6N first integrals are needed. But as it turned out we are able to find only ten first integrals, namely: energy integral, three turning moment integrals that traditionally are called area integrals and six location and velocity of the N- body system center of mass integrals. Thus , there is a lack of (6N-10) integrals to obtain general solution of N-body motion problem with intrinsic gravity. Notice that the same situation takes place with General Theory of Relativity. Early studies revolved around the Hamilton variational principle of least action and evidently incompleteness of modern astronomical mechanics directly connected with limitations of the Hamilton principle.

There are no immovable and motionless bodies and using results of our research it may be supposed that rotational material is the owner of gravity and it is intrinsically necessary for micro and macro bodies to possess the turning moment. Thus rotation of bodies is essential property of gravity.